\documentclass[12pt]{article}

\usepackage{newtxtext,newtxmath}

\usepackage{graphicx}

\usepackage[letterpaper,margin=1in]{geometry}

\renewenvironment{abstract}
	{\quotation}
	{\endquotation}

\date{}

\makeatletter
\renewcommand{\fnum@figure}{\textbf{Figure \thefigure}}
\renewcommand{\fnum@table}{\textbf{Table \thetable}}
\makeatother

\usepackage{scicite}

\usepackage{url}

\def\scititle{
	The Origin of Sound Damping in Amorphous Solids: Defects and Beyond
}
\title{\bfseries \boldmath \scititle}

\author{
	Elijah Flenner$^{1\ast}$, 
	Grzegorz Szamel$^{1\ast}$\and
	\small$^{1}$Chemistry Department, Colorado State University, Fort Collins \& 80523, USA.\and
	\small$^\ast$Corresponding author. Email: flennere@gmail.com\and
	\small$^\ast$Corresponding author. Email: Grzegorz.Szamel@colostate.edu
}

\begin{document} 

\maketitle

\begin{abstract} \bfseries \boldmath
 Comprehending sound damping is integral to understanding the
  anomalous low temperature properties of glasses. After decades of
  studies, Rayleigh scaling of the sound attenuation coefficient
  with frequency, $\Gamma \propto \omega^{d+1}$, became generally accepted.
  Rayleigh scaling invokes a picture of scattering from defects. It
  is unclear how to define glass defects.
  Here we use a particle level contribution to sound damping
  to determine areas in the glass that contribute more
  to sound damping than other areas,
  which allows us to define defects.
  Over a range of stability, sound damping scales linearly with the fraction of particles in the defects.
  However, sound is still attenuated in ultra-stable glasses where no defects are identified.
  We show that sound damping in these glasses is due to nearly uniformly distributed non-affine forces
  that arise after macroscopic deformation.
  To fully understand sound attenuation in glasses one has to consider contributions
  from defects and a defect-free background, which represents a different paradigm of sound damping in glasses.
\end{abstract}

\noindent
{\Large\textbf{Teaser}}\\
Sound attenuation in glasses exhibits features associated with scattering from defects without the presence of defects.

\noindent

\section*{Introduction}
The disordered structure of glasses results in low temperature properties that are different from those of 
crystals, and an explanation
of this finding 
remains an extensive area of research \cite{Ramos2023,Zeller1971}.  
It was initially thought that low frequency vibrations of glasses would be the plane waves
of the Debye model, as they are in crystals. 
The seminal paper by Zeller and Pohl \cite{Zeller1971} showed that the low-temperature 
specific heat and thermal conductivity of glasses were inconsistent with 
the Debye model. Later, inelastic neutron scattering \cite{Buchenau1984} and Raman 
scattering \cite{Schroeder2004,Sokolov1993} showed that there is an excess in the density of 
states above the Debye prediction. 

The excess vibrational density of states results in a peak in 
the reduced specific heat $C_p/T^3$ at a temperature between 3K and 15K \cite{Ramos2023,Talon2002}. 
In the same temperature range a plateau in the thermal conductivity appears \cite{Ramos2023,Zeller1971}, which suggests 
a connection between the two observations. 
In order to fit both the reduced specific heat peak and the thermal conductivity 
plateau with the same parameters, Yu and Freeman \cite{Yu1987} found that they needed 
$\omega^4$ Rayleigh scaling of sound damping. 

Rayleigh scaling invokes a picture of scattering from defects, but it is an ongoing question as to what constitutes a 
defect in a glass. 
Previous to Yu and Freeman, Zaitlin and Anderson  \cite{Zaitlin1975} considered density fluctuations as defects, but 
were unable to find reasonable values for Rayleigh scattering coefficients to quantitatively reproduce the thermal 
conductivity plateau.
This observation leads to the question of the identity of the 
defects that are responsible for Rayleigh scaling. 

A possible definition of a defect comes from the soft potential model \cite{Ramos2023}.
The soft potential model postulates the existence of 
excitations that couple to sound waves resulting in sound damping 
that scales as $\omega^4$ \cite{Buchenau1972,Schober2011}.  
Wang \textit{et al.}\ \cite{Wang2019a} found that sound damping scales linearly with the density of quasi-localized vibrational modes,
defined through the vibrational mode participation ratio \cite{Wang2019,Mizuno2017},
suggesting that quasi-localized modes may be scattering defects. However, they did not establish a direct
connection between the quasi-localized modes and sound damping, and the question still remained if the 
quasi-localized modes should be considered damping defects.

Some theories do not explicitly invoke defects and also predict Rayleigh scaling. Fluctuating elasticity theory describes glasses as elastic
materials with spatially varying elastic constants. When the spatial extent of elasticity fluctuations is much smaller than the wavelength of 
the sound wave, fluctuating elasticity theory predicts Rayleigh scaling of sound damping.  
Since elasticity is a property of a bulk material, it is not immediately clear how to define it locally \cite{Mizuno2013}.
For two-dimensional systems Kapteijns \textit{et al.} \cite{Kapteijns2021} argued that sample-to-sample elasticity fluctuations can be used to 
determine the relative scaling of sound attenuation, but this still leaves the question of the identification of elastic defects associated with sound damping. 

Mahajan and Pica Ciamarra \cite{Mahajan2021} 
examined this question. They found that local elastic constants defined in a specific way 
(see Supplemental Material of Ref.~\cite{Mahajan2021})
have fluctuations that have the same dependence on a glass model's potential cutoff  as the sample-to-sample fluctuations examined 
in Ref.~\cite{Kapteijns2021}.
They defined a coarse graining length that they identified with the characteristic defect size. 
They used this identification to show that correlated fluctuating elasticity theory \cite{Schirmacher2010,Schirmacher2011,Marruzzo2013}, 
with one adjustable parameter, accurately captures the relative dependence of sound damping on glass properties.

Rayleigh scaling is also predicted by theories where there is no clear way to define defects. Euclidean random matrix model,
which posits that an amorphous solid may be approximated as a set of randomly placed sites connected by harmonic springs,
leads to Rayleigh scaling of sound damping \cite{Ganter2010,Fuchs2024,Vogel2023}. 
In this model
defects cannot be identified beyond individual sites. 

Baggioli and Zaccone \cite{Baggioli2022} developed an approximate theory that predicts Rayleigh scaling of sound damping without invoking the concept of
defects. Their theory invokes an averaging procedure that smears out any explicit effect of defects.

All of the theories mentioned above provide a reasonable starting point and predict Rayleigh scaling 
of sound damping in three dimensions. 
However, they leave a confusing picture to the underlying damping mechanism, and if defects
are necessary for Rayleigh scaling. To differentiate the different pictures, one needs 
a quantitatively accurate theory with no fitting parameters. Only then one can determine the role of defects in sound damping. 

Here we  use a recently developed microscopic theory \cite{Szamel2022} that accurately predicts sound damping in the
harmonic approximation with no adjustable parameters.   Within many glasses we are able to find areas that 
result in strong sound damping over the frequency range where we observe Rayleigh scaling. We identify these areas as sound damping defects.
We find that the fraction of particles in these defects scales linearly with the Rayleigh scaling coefficient for a series of glasses with widely different stabilities. 
However, we find finite sound damping and Rayleigh scaling for exceptionally well annealed glasses where we do not  find any defects,
and thus the defects cannot be the sole source of sound damping. 

Sound damping 
without defects  originates from small non-affine motions that occur due to displacements induced by the
sound wave. These motions have to be present for any non centro-symmetric structure. The resulting picture is that sound damping has a defect
contribution on top of a defect free background, which represents a different paradigm in the understanding of sound damping in 
glasses. Our theory quantitatively captures both effects. 



\section*{Results}

 \subsection*{Theory and Simulation Comparison}
 
 We study a two-dimensional glass forming polydisperse mixture of spheres interacting 
 via a $r^{-12}$ repulsive potential in a fixed volume $V$. See Materials and Methods
 for details.
 Using the swap algorithm \cite{Ninarello2017}, our model system can be equilibrated down to temperatures below
 the mode coupling temperature of 0.123 and the estimated glass transition temperature of 0.082.
 Glasses of different stability are created by equilibrating a fluid 
 and quenching it into an inherent structure. These glasses are labeled by the temperature from which they were quenched, which we refer to as the parent temperature $T_p$.
 The parent temperature is similar to the fictive temperature, and the lower the $T_p$ the more stable the glass.


 To calculate sound damping in simulations and as a starting point of our theory, we consider a glass
 undergoing harmonic vibrations around an inherent structure. The
 equations of motion read
 \begin{equation}
 \label{eqmotion}
 \partial_t^2 \mathbf{u}_n = \ddot{\mathbf{u}}_n = - \sum_m \mathcal{H}_{nm} \cdot \mathbf{u}_m,
 \end{equation}
 where $\mathcal{H}$ is the Hessian calculated at the inherent structure positions $\{ \mathbf{R}_n \}$,
 $\mathbf{u}_n$ is the displacement of particle $n$ from $\mathbf{R}_n$. More details on the sound 
 damping theory and details of the calculation is found in Materials and Methods.
 The theory predicts that the attenuation coefficient of a transverse sound wave in the low frequency limit can be expressed as
  \begin{eqnarray}
   \Gamma_T(\omega) &=& k^2 \sum_{\omega_p} \delta(\omega-\omega_p) \frac{\pi}{2 \omega_p^2} \left<1\right|\mathcal{X} \left| \mathcal{E}(\omega_p) \right>^2 
   \nonumber \\
   \label{self}
   &=& (\omega/v_T)^2 \sum_{\omega_p} \delta(\omega - \omega_p) \epsilon(\omega_p) .
 \end{eqnarray}
 In Eq.~\eqref{self}
 $\left< 1 \right| \mathcal{X}$ is the non-affine force field due to a deformation,  
 $\left| \mathcal{E}(\omega_p)\right>$ is an eigenvector of the Hessian $\mathcal{H}$ corresponding to eigenfrequency $\omega_p$, and
 $v_T$ is the transverse speed of sound.
 Details of the calculation for finite systems can be found in Materials and Methods. 
 We note that inner product $\left<1\right|\mathcal{X} \left| \mathcal{E}(\omega_p) \right>^2$ features in the 
 theory of Baggioli and Zaccone \cite{Baggioli2022}, which was developed independently at the same time as our theory.
 However, the complete Baggioli and Zaccone's  expression for sound damping differs from ours. 
 
 We find that sound damping calculated from simulations (squares) and theory (dashed lines) agree very well
 in the $\omega^3$ (dotted line) Rayleigh scaling region, Fig.~\ref{damp}. We emphasize that
 our theory has no adjustable parameters, and thus it captures the relative change as well
 as the magnitude of sound damping. 
 We compare the results of the calculation using Eq.~\ref{self} to fits of the    
 function $\Gamma_T(\omega) = B_3 \omega^3$ 
 in Table~\ref{B3T}. 
 
To examine the properties of the glass that give rise to sound damping, we calculate the mode level 
contribution to sound damping using $\epsilon(\omega_p)$.  
 Shown in Fig.~\ref{cont} are the values of $\epsilon(\omega_p)$ calculated at each parent temperature for 40 glass samples
 for the range of $\omega_p$ needed to determine sound damping for the smallest wavelength sound wave 
 allowed due to periodic boundary conditions.  We find a dramatic change of the mode 
 level contribution with increasing stability, i.e.\ decreasing $T_p$.
 

 
For $T_p = 0.2$, Fig.~\ref{cont}\textbf{(A)}, there are clusters around frequencies corresponding to the first two
transverse sound waves, but in addition there are contributions at frequencies between these clusters. The 
value of $\epsilon(\omega_p)$ varies by several orders of magnitude for a given $T_p$. The overall scale of  $\epsilon(\omega_p)$ decreases
with decreasing $T_p$; it is much larger for $T_p = 0.2$ than for $T_p = 0.101$ and 0.03. 

For $T_p = 0.101$, Fig.~\ref{cont}\textbf{(B)}, there are two distinct clusters of modes with three modes making a noticeably 
larger contribution than other modes. If the Debye model was an accurate description of the low energy excitations, each 
configuration should contribute 4 modes to the first transverse wave and 4 modes to the second transverse wave, 
which represents the two clusters in Fig.~\ref{cont}\textbf{(B)}. Since we used 40 configurations for the calculation, according to the Debye model 
the first cluster should contain 160 modes. However, there are 161 modes in the highlighted box. Therefore, there is
one discrete excess mode over the Debye model in this frequency range. We note that no other modes contribute to sound damping
for the lowest frequency sound wave than the modes shown in Fig.~\ref{cont}.

For $T_p = 0.03$, Fig.~\ref{cont}\textbf{(C)}, there are two distinct clusters with 160 modes as expected for the Debye model. For the
40 configurations we studied, there are no excess modes obtained from diagonalizing the Hessian in this frequency range. 
Additionally, $\epsilon(\omega_p)$ is much smaller than 
for $T_p = 0.2$.
In Section Damping Defects we study what properties of the eigenvectors give rise to such large differences in $\epsilon(\omega_p)$.

\subsection*{Damping Defects}
\label{defects}
 
 To examine if there are areas of the glass that contribute to sound damping more than others, we 
 examine a particle level contribution to damping. 
 To this end we write $\sqrt{N}\left<1\right| \mathcal{X}$ in terms of the contributions of individual particles, which we denote $\Xi_n^\alpha$,
\begin{equation}
  \sqrt{N}\left<1\right| \mathcal{X} = (\Xi_1^x, \Xi_1^y, \ldots, \Xi_n^x, \Xi_n^y, \ldots, \Xi_N^x, \Xi_N^y).
  \end{equation}
The low-frequency sound damping is proportional to $\{\sum_n [\Xi_n^x E_n^x(\omega_p) + \Xi_n^y E_n^y(\omega_p)]\}^2$, where $E_n^{x,y}(\omega_p)$ 
are the components of the eigenvector corresponding to frequency $\omega_p$. We define the particle level contribution
\begin{equation}
  C_n = \Xi_n^x E_n^x(\omega_p) + \Xi_n^y E_n^y(\omega_p).
\end{equation}
The quantity $C_n$ can be positive or negative and sound damping is given by the square of the sum of these contributions.
Our hypothesis is that there exists spatial regions with large $|C_n|$ and these regions make a relatively large contribution 
to sound damping.  

Two obvious reasons that could make  $|C_n|$ large is that the magnitude of the vector $\mathbf{\Xi}_n = (\Xi_n^x, \Xi_n^y)$ is large
or the magnitude of the vector $\mathbf{E}_n(\omega_p) = [E_n^x(\omega_p),E_n^y(\omega_p)]$ is large.  
We will show that the eigenvectors for which there exist clusters of particles 
with large $|\mathbf{E}_n(\omega_p)|$ have a large contribution to sound damping 
over a range of frequencies. We classify these particles as belonging to defects.  
We find that these defects  can strongly influence sound damping, but they are not necessary for 
sound damping in glasses. Importantly, Rayleigh scaling of sound damping can occur without defects. 


It is instructive to examine the contributions to sound damping for our $T_p = 0.101$ glass, Fig.~\ref{cont}\textbf{(B)},
since there are two easy to identify modes that makes a large contribution to sound damping. 
We denote these mode as m1 and m2 in Fig.~\ref{cont}\textbf{(B)}.  Both of these modes originate from the same configuration.

Shown in Fig.~\ref{vis2}\textbf{(A)} is the non-affine force field due to a simple
shear deformation for the configuration with the eigenvector with the largest 
contribution to sound damping, Fig.~\ref{vis2}{(b), labeled m1 in Fig.~\ref{cont}\textbf{(B)}.
There is no obvious regions of large $|\mathbf{\Xi}_n|$.  
In contrast, there is a
cluster of large $|\mathbf{E}_n|$ in the eigenvector m1. 

Shown in Fig.~\ref{vis2}\textbf{(C)} 
is a color map of $|C_n|$ indicating the size of the contribution to sound damping for each particle corresponding to 
the mode shown in Fig.~\ref{vis2}(b). There is correspondence with the largest values of $|C_n|$
with the largest $|\mathbf{E}_n(\omega_p)|$.  In Fig.~\ref{vis2}\textbf{(D)} we show the particle level contributions to damping within the mode labeled 
m2 in Fig.~\ref{cont}\textbf{(B)}, which has the second largest contribution to the lowest frequency sound wave and comes from the
same configuration as m1. In the same region as for m1, the particle level contribution to sound damping is the largest.
 
One region of space can make a large contribution to sound damping over a range of frequencies. These regions 
can be considered defects with regards to sound wave propagation. Our theory predicts that these defects
are large sources to sound damping. 
However, since the non-affine force field is non-zero everywhere in the glass, defect free areas also contribute to sound damping. 

To determine areas of the glass where there is strong damping, we want to find areas where
$|\mathbf{E}_n(\omega_p)|$ is larger than expected for a plane wave over a range of frequencies. If 
the eigenvector $\left| \mathcal{E}(\omega_p)\right>$ is a plane wave, then $|\mathbf{E}_n(\omega_p)|^2 \le 2/N$ for 
each particle. Rather than focusing on a single frequency $\omega_p$ we consider the range of frequencies that encompasses
the Rayleigh scaling regime for every $T_p$. For our system size, this results in the 24 lowest frequency eigenvectors 
(excluding the uniform translations). 
Thus, we consider a particle part of a defect if $S_n = (N/2) \sum_{p=1}^{24} |\mathbf{E}_n(\omega_p)|^2 /24 > 1$. We 
define $w_n = 1$ if a particle is within a defect and zero otherwise. 

Shown in Fig.~\ref{vis2}\textbf{(D)} is a color map of $S_n$ where
the red regions correspond to defects in the configuration with the eigenvectors corresponding to m1 and m2 in Fig.~\ref{cont}\textbf{(B)}.
The procedure clearly picks out the region of large damping seen in Fig.~\ref{vis2}(b), 
Fig.~\ref{vis2}\textbf{(C)}, and Fig.~\ref{vis2}\textbf{(D)} as well as some smaller regions.  

For each parent temperature we can determine the glass configuration with the largest number of particles 
within defects and the least number of particles within defects. For our $T_p = 0.2$ 
glasses we identified a defect within every glass configuration. Shown in Fig.~\ref{tpdefect}\textbf{(A)} is 
a color map of $S_n$ of the configuration with the largest number of particles within defects. 
Figure \ref{tpdefect}\textbf{(B)} shows a configuration with an average number of particles within defects,
and Fig.~\ref{tpdefect}\textbf{(C)} shows the configuration with the smallest number of particles within defects for the $T_p = 0.2$ glasses.  


For our $T_p = 0.101$ glasses defects can be found in 37 out of the 40 glass configurations. Shown in  
Fig.~\ref{tpdefect}(\textbf{D}) is a color map of $S_n$ for the configuration with the largest number of particles within defects, 
the configuration with about the average number of particles in defects is shown in Fig.~\ref{tpdefect}\textbf{(E)}, and
a configuration with the smallest number of particles in defect (zero) is shown in Fig.~\ref{tpdefect}\textbf{(F)}. Visually
we can see a large change in the number and size of the defects with increasing stability.
For our $T_p = 0.03$ glasses we did not find defects in any of the 40 glass configurations. 

To further motivate our definition of a defect we plot the Rayleigh scaling coefficient $B_3$ found 
from fits of $\Gamma(\omega) = B_3 \omega^3$ versus the average fraction of particles within defects, $c = \left<\sum_n w_n\right>/N$,  
in Fig.~\ref{defect}\textbf{(A)}. We find that $B_3$ increases linearly with the density of particles in a defect for $T_p \ge 0.085$, but the $y$-intercept is non-zero.
Therefore, the defects can be a large contribution, but are not the sole contribution to sound damping.  

While we are only interested in defects that influence sound damping here, we note that our definition of a defect bears resemblance to 
defects defined by other researchers. Widmer-Cooper \textit{et al.} showed 
that particles with a large participation fraction, defined as the sum of $|\mathbf{E}(\omega_p)|^2$ over the 30 lowest-frequency modes, 
correlated with particles that are most likely to rearrange in a supercooled fluid \cite{WidmerCooper2008}.  Our defects would also be areas of
large participation fraction. 
A related quantity, the vibrality $\Psi_n = \sum_p |\mathbf{E}_n(\omega_p)|^2/\omega_p^2$,
was found to be a good indicator of a structural defect responsible for plastic flow \cite{Richard2020}.  These studies and ours suggest that
defects in glasses can be identified through $|\mathbf{E}_n(\omega_p)|^2$ for the low frequency modes. 

\subsection*{Damping Without Defects}

To estimate the contribution to damping due to areas without defects,
we calculate sound attenuation in the Rayleigh regime using the plane waves of the Debye model instead of the
eigenvectors of the Hessian matrix. For this calculation we use the same non-affine force field and we do not change
the frequencies corresponding to the eigenvectors. 
We are interested in the effects of changing the eigenvector structure alone, which removes our defects. 

Shown in Fig.~\ref{defect}\textbf{(B)} is the Rayleigh scaling coefficient obtained from the fits to the full theory, $B_3$ (black circles),
and the coefficient obtained from the calculation that uses plane waves instead of the actual eigenvectors, $B_3^{\text{pw}}$ (red squares). 
For $T_p = 0.03$ this procedure gives $B_3^{\text{pw}} = 0.018\pm0.002$, which is statistically the same as the one calculated using the 
eigenvectors of the Hessian. Therefore, defects do not play a role in 
sound damping for our $T_p = 0.03$ glasses.  The increase in $B_3^{\text{pw}}$ with increasing $T_p$ is due to an increase in the average magnitude of the 
non-affine forces with increasing $T_p$. 
With decreasing stability the plane wave approximation becomes less accurate, and largely underestimates
sound damping in our poorly annealed glass. 
   

The resulting picture is that non-affine forces, \textit{i.e.} non-zero values of $\Xi_n$, are important in the understanding of sound damping in the harmonic approximation. 
The magnitude of sound attenuation is set by two contributions, one coming from defects and another one that originating from defect-free
areas. The defect contribution dominates for moderately to poorly annealed glasses.

The fraction of particles in the defects depends on the glass stability, \textit{i.e.} on $T_p$.
We found that the dependence of $c$ on $T_p$ can be described reasonably well by
$c \propto e^{-b/(T_p-T_0)}$ over the full $T_P$ range, inset to Fig.~\ref{defect}\textbf{(B)}.  We find that $T_0 = 0.0195 \pm 0.005$, 
which is consistent with the defect density going to zero around $T_p = 0.02$. We find that 
Boltzmann-like scaling $c \propto e^{-b/T_p}$ also provides an accurate description of the 
data. We note that Boltzmann-like scaling involving
the parent temperature was observed for the density of quasi-localized vibrational excitations \cite{Rainone2020}.
Boltzmann-like scaling involving an effective temperature was derived for the density of shear transformation zones \cite{Bouchbinder2009}.
While both quasi-localized excitations and shear transformation zones address similar physics as our defects, a precise relationship
between them 
is left for future study.

To rationalize how Rayleigh scaling occurs without defects, we refer to the result of Zaccone and Scossa-Romano \cite{Zaccone2011}
who found that $\left<1|\mathcal{X}|\mathcal{E}(\omega_p)\right>^2$ scales as $\omega_p^2$ within the isotropic approximation for the Hessian.
We numerically verified that this is approximately accurate.
Therefore, we expect that $\Gamma_T(\omega) \approx (A_D \bar{\epsilon}/v_T^2) \omega^3 $, where
$A_D$ is the Debye level, and $\bar{\epsilon}$ is an average $\epsilon(\omega_p)$, see Methods. 
We find that this approximation reproduces $B_3$, see Table~\ref{B3T}, but the uncertainty is larger. 
Thus, the existence of defects is not the necessary condition for Rayleigh scaling, and approximate sound damping can 
be determined from approximations to $\bar{\epsilon}$. 
Future theories 
should provide approximations to $\bar{\epsilon}$ that can be obtained from experimentally measurable 
quantities. 

\section*{Discussion}

While we are able to determine the location of damping defects, we have not determined the relationship of these
defects with other theories. 
Defects shown in 
Fig.~\ref{cont} resemble low-frequency quasi-localized excitations \cite{Gartner2016,Wijtmans2017,Kapteijns2020}, which lead to 
sound damping in the soft-potential model \cite{Buchenau1972}.  
Using an approximation where we replace quasi-localized modes hybridized with phonons with localized modes results in sound damping scaling as the density of quasi-localized 
excitations $g_s(\omega)$.  Buchenau \textit{et al.} \cite{Buchenau1972}
derived an expression that is proportional to $g_s(\omega)$ for the soft potential model, and thus would bear some resemblance to our theory using this approximation. 
Further work is needed to explore connections between the two theories. In particular, it is not clear how defect-free sound attenuation
can be described within the soft potential model. 

While we used properties of the eigenvectors to find defects, recent work suggests that defects may be found by examining the non-affine displacement
field. It was shown in Ref.~\cite{Baggioli2021} that dislocation-like topological defects associated with plastic yielding can be identified in the non-affine displacement field
for a two-dimensional glass.  
Recent work has expanded on the characterization of these defects and their relationship to plastic failure \cite{Desmarchelier2024,Bera2024}. Future work should examine if these 
dislocation-like topological defects are also associated 
with areas of strong sound damping. 

Kapteijns \textit{et al.}\ \cite{Kapteijns2021} dramatically reduced sound damping by reducing internal stresses, which in turns introduces a gap
in the low frequency spectrum of quasi-localized excitations \cite{Lerner2018}. This may also remove defects. They found sample-to-sample
elasticity fluctuations could describe the relative change of sound damping. These sample-to-sample elasticity fluctuations may be related to changes in the non-affine force field, 
which controls defect free sound damping.  

Non-affine forces play a role both in our microscopic theory of sound attenuation, and in definitions of local elastic constants \cite{Mizuno2013}.
Further investigation of local elastic constants may help to establish connection between our theory and the fluctuating elasticity theory.
Extending Mahajan \textit{et al.}'s \cite{Mahajan2021} comparison between sound attenuation
and elastic constants fluctuations to a wider variety of systems may help clarify this issue. 

Mahajan and Pica Ciamarra \cite{Mahajan2024} studied the spatiotemporal pattern of sound damping in simulations of model three dimensional glasses. They 
measured a particle level attenuation and found that areas of largest attenuation were correlated with quasi-localized modes. Future work should examine 
the spatiotemporal character of sound damping in glasses where we do not observe defects.  

\section*{Materials and Methods}

\subsection*{Simulations} \label{simulation} 

 We study a system of $N$ polydisperse particles confined to a two-dimensional volume $V = L^2$ with 
 $r^{-12}$ repulsive interactions that is cutoff and shifted 
 so that the potential and its derivatives are continuous up to the second derivative. 
 The same system was studied by Berthier \textit{et al.}\ \cite{Berhtier2019}.
 The interaction potential is given by 
 \begin{equation}
 v_{ij} = v_o \left( \frac{\sigma_{ij}}{r} \right)^{12} + c_o + c_1 \left( \frac{r}{\sigma_{ij}}\right)^2 + c_2 \left( \frac{r}{\sigma_{ij}} \right)^4,
 \end{equation}
 where $\sigma_{ij} = 0.5(\sigma_i + \sigma_j)(1-0.2|\sigma_i-\sigma_j|)$.
 The potential parameter $v_0$ sets the units of energy and we set Botzmann's constant equal to one. The 
 diameters of the particles $\sigma_i$ are randomly drawn from a distribution of the form $f(\sigma) = A \sigma^{-3}$ 
 for $\sigma \in [\sigma_{\mathrm{min}},\sigma_{\mathrm{max}}]$ where $\sigma_{\mathrm{min}}/\sigma_{\mathrm{max}} = 0.45$. 
 The average diameter sets the unit of length. We cut and shift the potential at $1.25 \sigma_{ij}$. The results given here 
 are for systems of $N = 20,000$. 
 
 To create a glass we use configurations that were equilibrated at a 
 parent temperature $T_p$. We quench these configurations to a potential energy minimum using a 
 conjugate gradient algorithm in LAMMPS \cite{lammps}. The stability of the glass is determined by its parent temperature.
 We examine in detail three parent temperatures, a poorly annealed glass at $T_p = 0.2$, 
 an intermediate parent temperature $T_p = 0.101$, and a very stable glass $T_p = 0.03$.  
 
 \subsection*{Damping Theory} \label{theory}
 
 We consider the harmonic approximation where the equation of motion is
 \begin{equation}
 \label{motion}
 \partial_t^2 \mathbf{u}_n = \ddot{\mathbf{u}}_n = - \sum_m \mathcal{H}_{nm} \cdot \mathbf{u}_m,
 \end{equation}
 where $\mathcal{H}_{nm}$ is the Hessian calculated at the inherent structure positions $\{ \mathbf{R}_n \}$,
 $\mathbf{u}_n$ is the displacement of particle $n$ from $\mathbf{R}_n$.
 The initial conditions in the simulations are $\dot{\mathbf{u}}_n(t=0) = \mathbf{a} \sin(\mathbf{k} \cdot \mathbf{r}_n)$,
 $\mathbf{u}_n = \mathbf{0}$,
 and $\mathbf{a} \cdot \mathbf{k} = 0$ for a transverse wave, and $\mathbf{a}$ parallel to  $\mathbf{k}$ for a longitudinal wave. We determine sound damping
 in simulations by fitting the envelope of $C_{\mathbf{k}}(t) = [\dot{\mathbf{u}}(t)\cdot\dot{\mathbf{u}}(0)]/[\dot{\mathbf{u}}(0)\cdot\dot{\mathbf{u}}(0)]$ to $e^{-\Gamma t/2}$ 
 as was done in previous work \cite{Wang2019a}. 
 
 The theory \cite{Szamel2022} is formulated such that the initial conditions are $\mathbf{u}_n(t=0) = \mathbf{b} e^{-i \mathbf{k} \cdot \mathbf{R}_n}$ 
 and $\dot{{\mathbf{u}}}_n(t=0) =  0$.  We define a two-dimensional
 vector $\mathbf{e}_n$ such that $\mathbf{e}_n \cdot \mathbf{e}_n = 1$, and is identical for each $n$.
 Solving the equations of motion \eqref{motion}  is equivalent to 
 solving $\partial_t^2 \left| 1(t) \right> = -\mathcal{H}(\mathbf{k}) \left| 1(t) \right>$, where 
 $\mathcal{H}_{nm}(\mathbf{k}) = \mathcal{H}_{nm} e^{i \mathbf{k} \cdot(\mathbf{R}_n - \mathbf{R}_m)}$, 
 with the initial condition $\left| 1(t=0) \right> = \left| 1 \right> = N^{-1/2} (\mathbf{e}_1, \ldots , \mathbf{e}_N)$. 

 In practice, the low frequency limit of transverse sound damping $\Gamma_T(\omega)$ is calculated using the distribution \cite{Szamel2022}
  \begin{eqnarray}
 \label{selfd}
 \Gamma_T(\omega) &=& \frac{\omega^2}{v_T^2} \frac{1}{\delta \omega} \sum_{\omega_p \in \{\omega-\delta \omega/2,\omega+\delta \omega/2\}}  \epsilon(\omega_p),
 \end{eqnarray}
where $\epsilon(\omega_p) = (\pi)/(2 \omega_p^2) \left| \left<1\right| \mathcal{X} \left| \mathcal{E}(\omega_p) \right> \right|^2$
and $c_T$ is the transverse speed of sound. In practice one must include frequencies where plane-wave like modes exist in the
finite sized system.  
 In Eq.~\eqref{selfd} $\left| \mathcal{E}(\omega_p) \right>$ is a normalized eigenvector of $\mathcal{H}$ with eigenvalue $\omega_p^2$.
 When we refer to a mode or vibrational mode in this work, we are always referring to an eigenvector of the Hessian. 
 
 To determine $\mathcal{X}$ we first set $\mathbf{e}_n = (1,0)$, define the matrix $\mathcal{X}_{nm}^1 = \mathcal{H}_{nm}(Y_n - Y_m)$, and
 determine the distribution given in Eq.~\eqref{selfd}. We then set $\mathbf{e}_n = (0,1)$, define the matrix $\mathcal{X}_{nm}^2 = \mathcal{H}_{nm}(X_n - X_m)$,
 and determine the distribution given by Eq.~\eqref{selfd}. We then average the distributions and fit the resulting distribution to
 $A \omega$ over a frequency range where Rayleigh scaling is 
 observed. The damping coefficient is given by $A \omega^3/v_T^2$. The results of these fits are given in Table \ref{B3T}. 
 We examine distributions with $\delta \omega$ ranging from 0.05 to 0.15. Since we are fitting a range of frequencies, we find 
 that the bin size makes little difference. 
 We find the speed of sound by using the theory given by Szamel and Flenner \cite{Szamel2022}.
  
 There are several important aspects to Eq.~\eqref{selfd}. First, it is a weighted distribution that would be proportional to the 
 density of states if the weights  $\epsilon(\omega_p)$ were all equal.  
 Therefore the density of states influences the frequency dependence of sound damping. 
 Second, the weights $\epsilon(\omega_p)$ can be thought of as different contributions from the vibrational mode
 $\left| \mathcal{E}(\omega_p) \right>$, and only modes around the frequency of the sound wave contribute to 
 sound damping. 
 
 While $\epsilon(\omega_p)$ differs by orders of magnitude depending on the details of the
 eigenvector and $\mathcal{X}$, on average it does not grow or decrease over the Rayleigh scaling
 regime, see Supplemental Material. Hence we can approximate the value of the distribution 
 by calculating $\bar{\epsilon} = (2N/\mathbb{N}_p) \sum_p^{\mathbb{N}_p} \epsilon(\omega_p)$
 where the sum is taken over the range of frequencies where Rayleigh scaling is observed. 
 For $T_p =0.2$ we took an average up to $\omega =0.24$, for $T_p = 0.101$ we averaged up 
 to $\omega = 0.55$, and for $T_p = 0.03$ we averaged up to $\omega = 1.4$. As long as we 
 included the first 24 modes per configuration in the average, the average varies by less than 20\%, which 
 is close to the uncertainty in all our calculations. 

\newpage


 \begin{figure}
 \includegraphics[width=0.9\columnwidth]{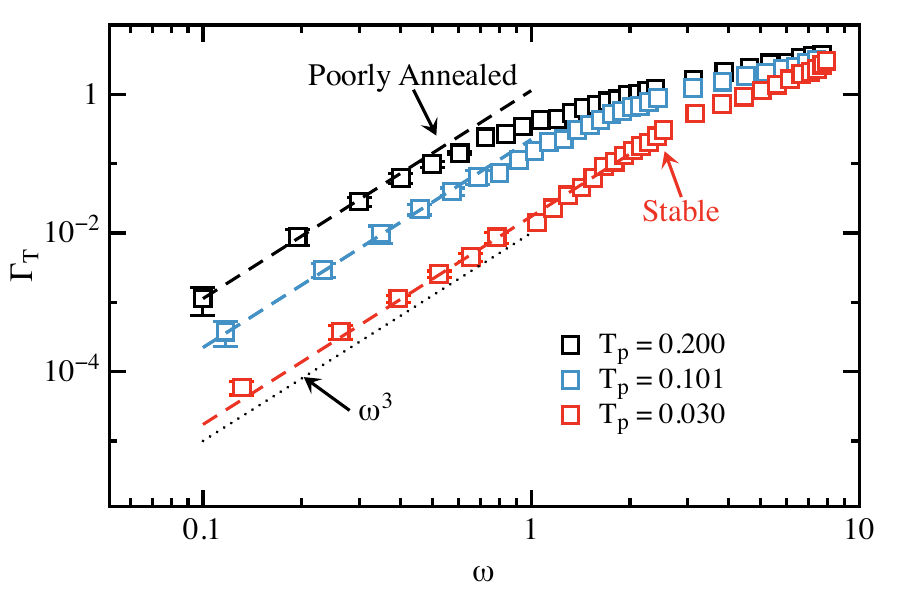}
 \caption{\label{damp}\textbf{Theory and simulation comparison.} Comparison of transverse sound damping $\Gamma_T$ calculated from simulations
 (squares) and the low wavevector expansion of the theory (dashed lines) as a function of frequency $\mathbf{\omega}$. 
 Glasses at three different stability are shown.
 A poorly annealed glass, $T_p = 0.2$ (black), a stable glass $T_p = 0.101$ (blue), and an 
 exceptionally stable glass $T_p = 0.03$ (red). There is near perfect agreement 
 between simulations and theory with no adjustable parameters. The dotted line
 represents Rayleigh scaling $\omega^3$, which is in good agreement with the results.}
 \end{figure}
 
 \begin{figure}
 \includegraphics[width=0.9\columnwidth]{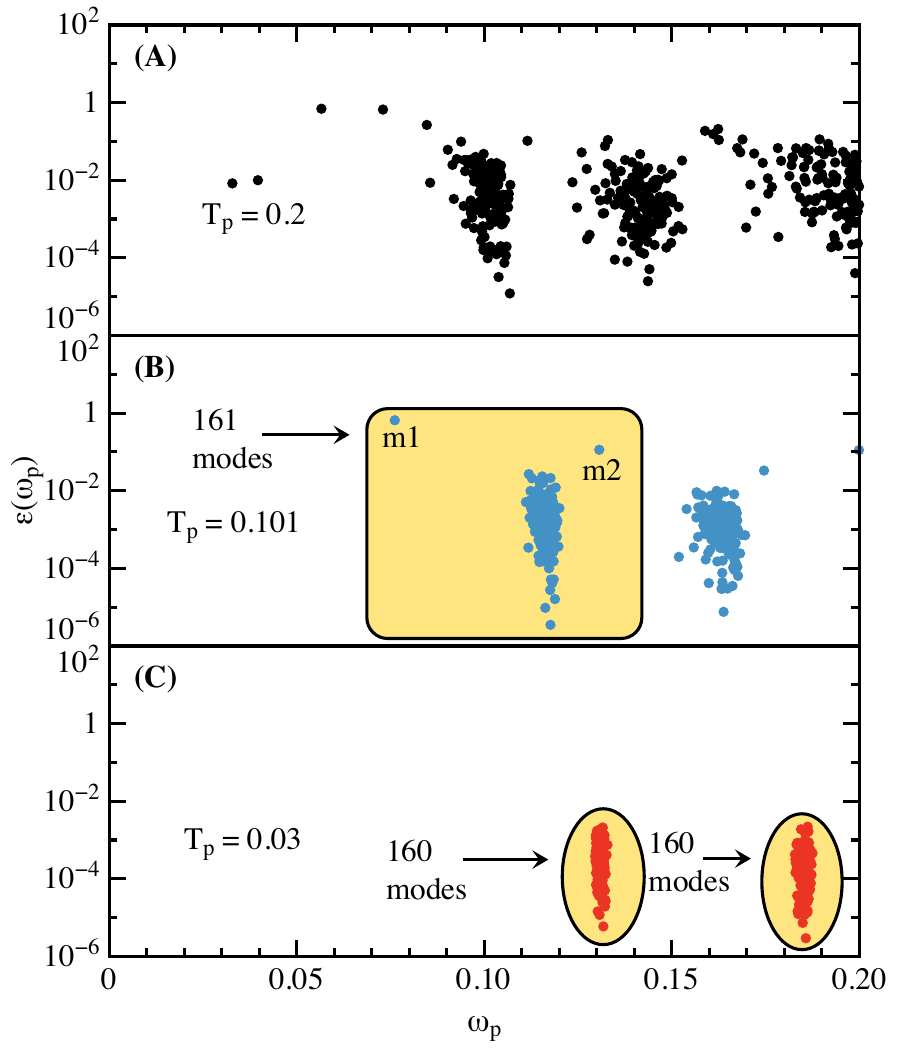}
 \caption{\label{cont}\textbf{Vibrational mode level contribution to damping.} The contribution to sound damping from different eigenvectors $\left| \mathcal{E}(\omega_p)\right>$
 with a frequency $\omega_p$. (\textbf{A}) A poorly annealed glass, $T_p = 0.2$. (\textbf{B}) A well annealed glass, $T_p = 0.101$. 
 (\textbf{C}) An exceptionally stable glass, $T_p = 0.03$.}
 \end{figure}
 
 \begin{figure}
\begin{center}
\includegraphics[width=0.95\columnwidth, bb = 0cm 11cm 27.9cm 19.05cm]{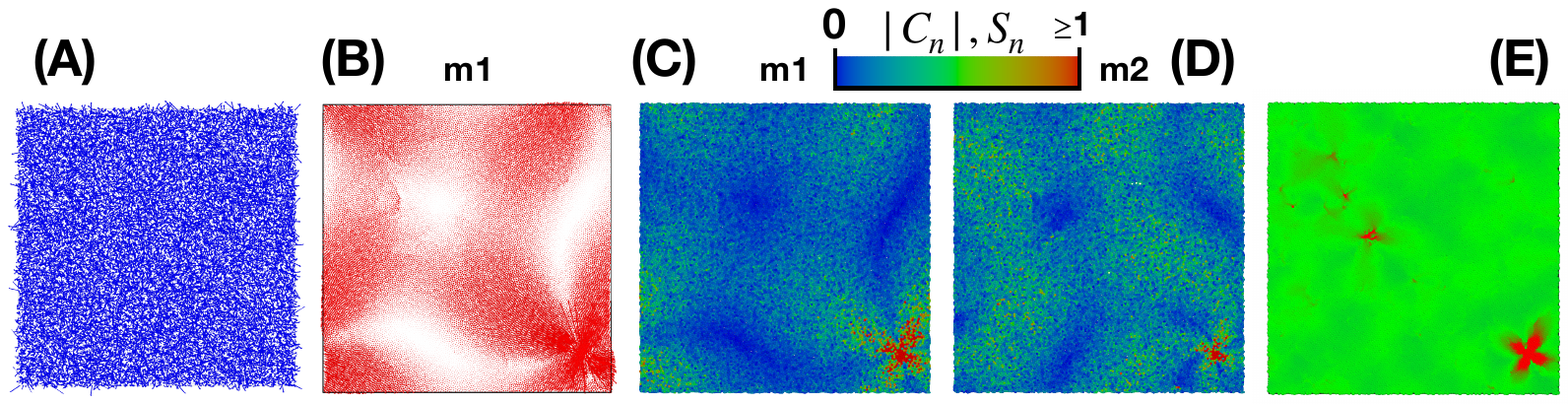}
\end{center}
\caption{\label{vis2}\textbf{Visualization of damping defects.}
(\textbf{A}) A non-affine force field for a shear deformation of the configuration represented in (\textbf{B})-(\textbf{E}).
(\textbf{B}) The eigenvector, m1 in Fig.~\ref{cont}\textbf{(B)}, with the largest contribution
to sound damping out of the forty $T_p=0.101$ glasses we studied. An area of 
large $|\mathbf{E}_n(\omega_p)|$  
can be found in the lower right corner. (\textbf{C}) Color map showing 
the relative contribution to sound damping $|C_n|$
for the same vibrational modes shown in (b), which shows that the particles within 
area of large $|\mathbf{E}_n(\omega_p)|$
make the largest contribution to sound damping for this vibrational mode. 
(\textbf{D}) The particle level contribution for the mode with the second largest contribution
to sound damping for $T_p = 0.101$ glasses (m2 in Fig.~\ref{cont}\textbf{B}), showing that the
same soft spot influences sound damping for both modes at different frequencies. 
This mode originates from the same glass as m1.
(\textbf{E}) A color map of $S_n$ using the same glass configuration as shown in (\textbf{A})-\textbf{(C)}. The defect particles identified using 
the procedure described in the text are colored red.}
\end{figure}

\begin{figure}
\includegraphics[width=0.95\columnwidth]{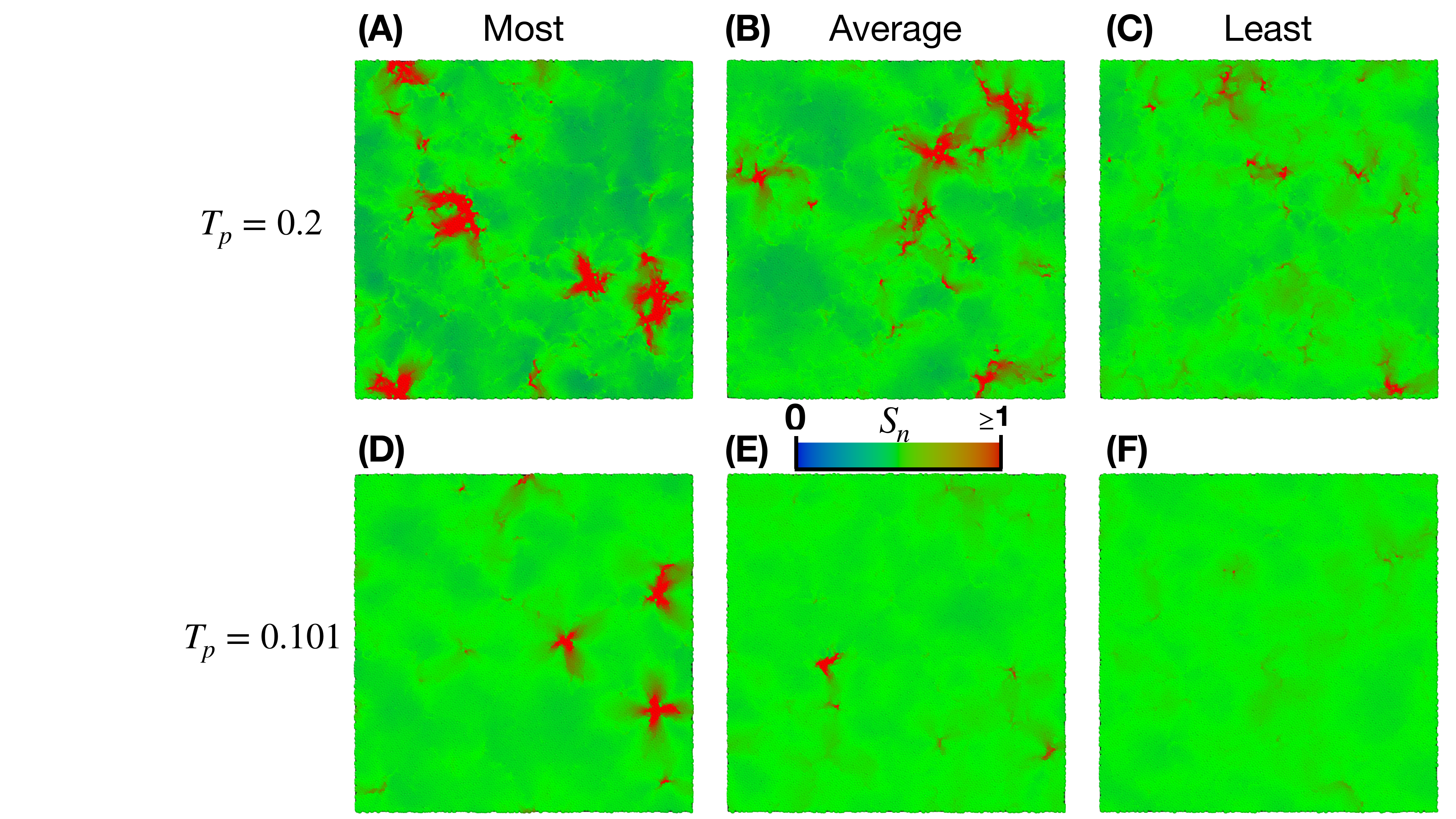}
\caption{\label{tpdefect} \textbf{Stability dependence of defects.} Stability and configuration dependence of defects. A color map of $S_n$ for
the $T_p = 0.2$ glass with \textbf{(A)} the most number of particles in a defect, \textbf{(B)} a glass with close to the average 
number of particles in a defect, and \textbf{(C)} the glass with the least number of particles in a defect.
 A color map of $S_n$ for
the $T_p = 0.101$ glass with \textbf{D} the most number of particles in a defect, \textbf{(E)} a glass with close to the average 
number of particles in a defect, and \textbf{(F)} a glass with the least number of particles in a defect. Three glasses had 
zero defects for $T_p = 0.101$. There were no defects for our $T_p = 0.03$ glasses.}
\end{figure}

\begin{figure}
\includegraphics[width=0.95\columnwidth]{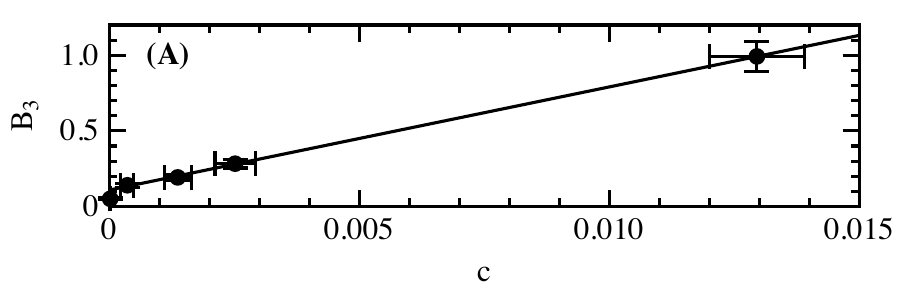}
\includegraphics[width=0.95\columnwidth]{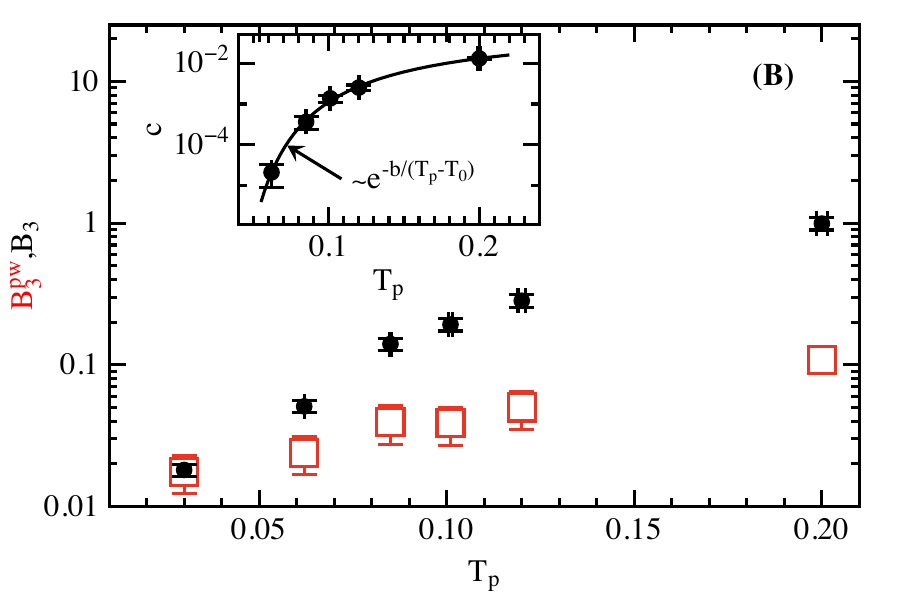}
\caption{\label{defect}\textbf{Defect contribution to damping.} \textbf{(A)} Strength of sound damping, $B_3$, as a function 
of the fraction of particles within a defect. The black line is a linear fit to $T_p \ge 0.085$. 
\textbf{(B)} Parent temperature $T_p$ dependence of sound damping. 
Black circles are $B_3$ obtained from fits of $\Gamma_T = B_3 \omega^2$. Red squares 
are $B_3^{\text{pw}}$ obtained from approximating the eigenvectors as plane waves, and represents an approximate 
contribution of defect free areas of the glass. The inset shows the 
$T_p$ dependence of the defect density, and the solid line is a fit to 
to  $c \sim e^{-b/(T_p-T_0)}$, where we find $T_0 = 0.0195 \pm 0.05$..}
\end{figure}

%


 \begin{table}
 \centering
\caption{\textbf{Damping coefficient determined using various methods.}
The first method, Fit, corresponds to $B_3$ obtained from fitting the squares shown in 
Fig.~\ref{damp} to $\Gamma_T(\omega) = B_3 \omega^3$. The second method uses the results obtained from Eq.~\eqref{self}. 
The third method uses $\epsilon(\omega)$ averaged over the Rayleigh scaling regime. }\label{B3T}%
\begin{tabular}{@{}llll@{}}
\\
\hline
$T_p$ & Fit, $B_3$ & Eq.~\eqref{self}  & $A_D \bar{\epsilon}/c_T^2$  \\
\hline
0.2    & $1.0\pm0.2$   & $1.2\pm0.2$  & $1.1\pm0.3$  \\
0.101    & $0.19\pm0.02$   & $0.21\pm0.02$  & $0.22\pm0.04$  \\
0.03    & $0.018\pm0.003$   & $0.015\pm0.002$  & $0.018\pm0.004$  \\
\hline
\end{tabular}
\end{table}

%


\clearpage 

%
\bibliography{science_template} 
\bibliographystyle{sciencemag}

\section*{Acknowledgments}
We thank Edan Lerner for a discussion on Rayleigh scattering in amorphous solids without defects. We also thank him and
 Massimo Pica Ciamarra for comments on the manuscript. 
\paragraph*{Funding:}
 EF and GZ gratefully acknowledge the support of NSF Grant No. CHE 2154241.
\paragraph*{Author contributions:}
Conceptualization: E.F. Methodology: E.F. Visualization: E.F. Supervision: G.S. Writing-original draft: E.F. Writing-review and editing: E.F. and G.S. 
\paragraph*{Competing interests:}
There are no competing interests to declare.
\paragraph*{Data and materials availability:}
The data presented in this publication is available at Dryad: https://doi.org/10.5061/dryad.cz8w9gjd8. 
All other data needed to evaluate the conclusions in this paper are present in the paper and/or the Supplementary Materials.




\newpage


\renewcommand{\thefigure}{S\arabic{figure}}
\renewcommand{\thetable}{S\arabic{table}}
\renewcommand{\theequation}{S\arabic{equation}}
\renewcommand{\thepage}{S\arabic{page}}
\setcounter{figure}{0}
\setcounter{table}{0}
\setcounter{equation}{0}
\setcounter{page}{1} 


%
%
%
%

\end{document}